\documentclass{PoS}

\title{UCIRC2: An Infrared Cloud Monitor for EUSO-SPB2}

\ShortTitle{UCIRC2}

\author{\speaker{Rebecca Diesing}, Khadijat Durojaiye, Seamus Flannery, Kameron Mehling, Noah Friedlander, Alexa Bukowski, Emily Donovan, Stephan Meyer, and Angela V. Olinto\\
       Department of Astronomy \& Astrophysics, KICP, EFI, The University of Chicago, USA\\}
       
\author{for the JEM-EUSO Collaboration \\}

\abstract{We describe the design and implementation of the University of Chicago Infrared Camera 2 (UCIRC2) built for monitoring cloud coverage during the EUSO-SPB2 flight (the second generation of the Extreme Universe Space Observatory on a Super Pressure Balloon). UCIRC2 uses two infrared (IR) cameras centered on 10$\mu$m and 12$\mu$m wavelengths to capture images of the clouds beneath EUSO-SPB2 in two bands spanning the thermal emission peak. Taken every minute, the IR images allow the determination of the height and coverage of clouds between the telescope and the ground. We discuss the design and construction of UCIRC2, including the techniques and design principles that make the module temperature and vacuum resilient. Additionally, we delineate the image reconstruction process and the pixel by pixel temperature calibration procedure. This paper will posit design and implementation suggestions for future ultra-high energy space telescopes.}

\FullConference{36th International Cosmic Ray Conference, ICRC2019\\
		24 July - 1 August, 2019\\
		Madison, Wisconsin, USA}

\begin{document}

\section{Introduction}

Ultra High Energy Cosmic Rays (UHECRs), cosmic rays (CRs) with energy above $10^{18}$ eV, are currently detected with Cherenkov water tanks and up-looking fluorescence detectors in observatories such as the Pierre Auger Observatory \cite{Auger} in Argentina and the Telescope Array \cite{TA} in Utah. In particular, UHECRs can be detected via a characteristic particle shower, called an Extensive Air-Shower (EAS), that occurs when an UHECR interacts with Earth's atmosphere. This EAS produces fluorescence of atmospheric nitrogen molecules, detectable in the 300-400 nm spectral band, as well as optical Cherenkov light. Because UHECRs are rare, ($<$ 1 per km${^2}$ per century), extremely large detector volumes are required to enable charged-particle astronomy. One way to increase detector volume is to observe the atmosphere from above. The second generation of the Extreme Universe Space Observatory on a Super Pressure Balloon (EUSO-SPB2) will do just that, making the first optical observations of ultra-high energy cosmic rays (UHECRs) by looking down upon the atmosphere. 

\begin{figure}
\begin{center}
\includegraphics[width=0.95\textwidth]{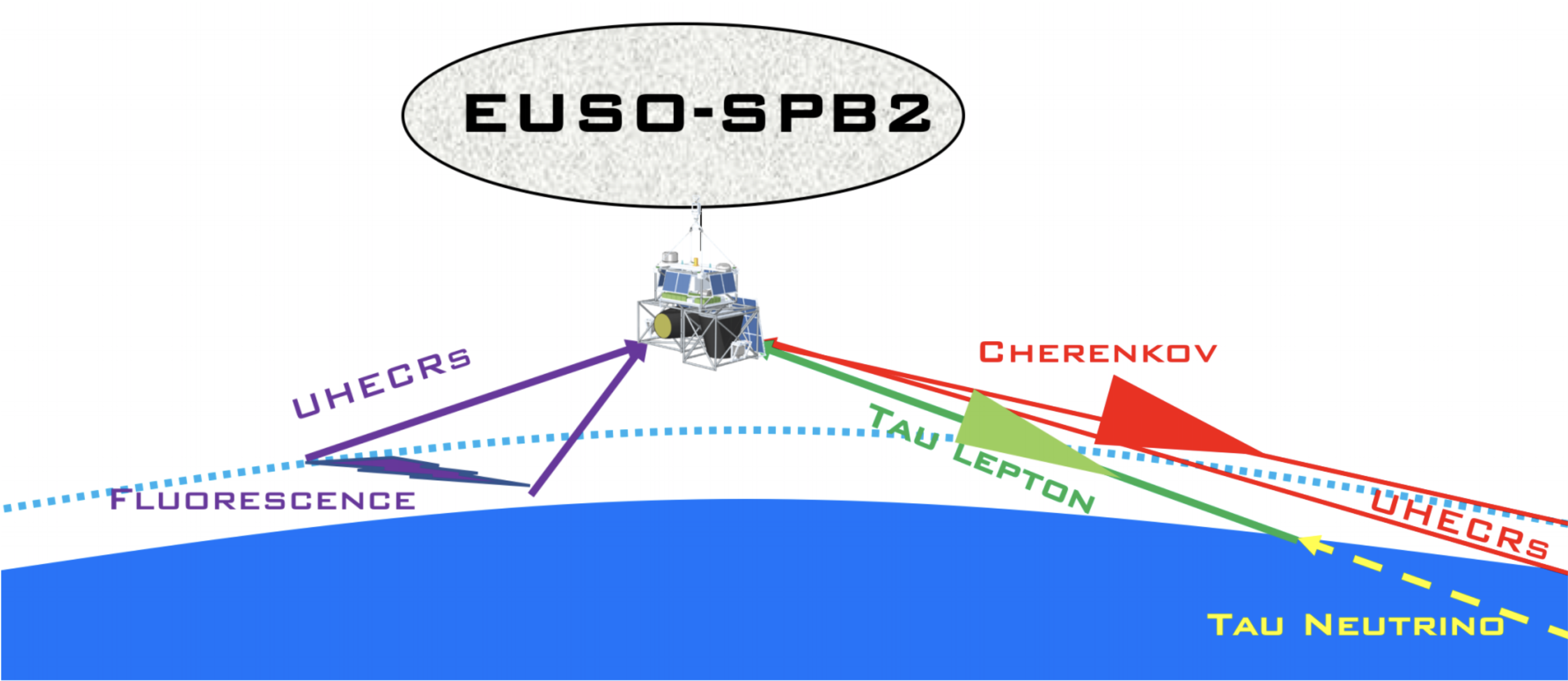}   
\end{center}
\caption{EUSO-SPB2's three detection modes: fluorescence from UHECRs (purple), Cherenkov from UHECRs (red), and Cherenkov from CNs (green).}
\label{fig:EUSOSPB2}
\end{figure}

A pathfinder to the more ambitious satellite mission POEMMA (Probe of Extreme Multi-Messenger Astrophysics), EUSO-SPB2 will detect UHECRs via two complementary techniques: looking down upon the atmosphere with a fluorescence telescope and looking towards the limb of the Earth to observe the Cherenkov signals produced by UHECRs above the limb. EUSO-SPB2 will also search for cosmic neutrinos signatures via the Cherenkov light from upward going tau leptons produced when a tau neutrino interacts near the surface of the Earth (see Figure \ref{fig:EUSOSPB2}) \cite{EUSOSPB2}. 

The presence of high clouds within the detectors' field of view (FoV) can significantly reduce the UHECR event detection rate and event energy calibration. Namely, it is possible for the peak of the EAS signal to occur beneath high clouds. Determining the exposure of EUSO-SPB to UHECRs requires knowledge of the effective detector volume, i.e., the volume of atmosphere within the field of view, above the clouds. Thus, EUSO-SPB2 requires continuous information about cloud coverage and height within the detectors' FoV. This is the responsibility of the second generation of the University of Chicago Infrared Camera (UCIRC2). 

The design and calibration of UCIRC2 improves upon that of UCIRC1, which flew onboard EUSO-SPB1. More information on UCIRC1 can be found in \cite{UCIRC1}.

\section{Method}

\begin{figure}
    \centering
    \includegraphics[width=0.95\textwidth]{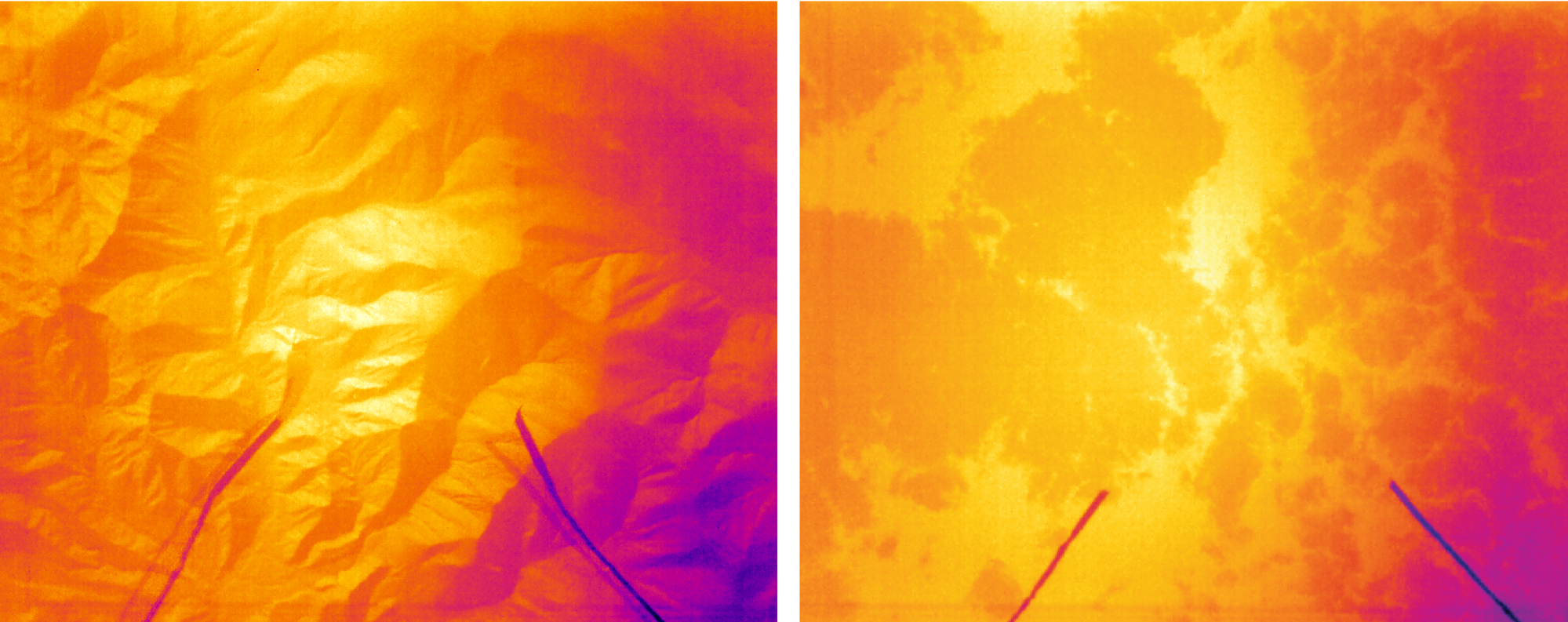}
    \caption{Uncalibrated images of mountains (left) and clouds (right) captured by UCIRC1, which flew on EUSO-SPB1 in 2017. Even without calibration, cloud coverage can be easily determined.}
    \label{fig:SamplePic}
\end{figure}

When EUSO-SPB2 is in observing (night) mode, IR images of the environmental conditions in and around the effective UHECR detection area are captured by UCIRC2 every 60 seconds. These images can be used to collect information about cloud coverage and altitude (cloud top height, CTH) within the field of view of the UHECR detectors (see Figure \ref{fig:SamplePic}).

By assuming that clouds are in thermal equilibrium with their surroundings, CTH can be inferred from cloud temperature, $T_{\rm c}$ which can be estimated using two brightness temperatures in bands near the cloud blackbody peak. More specifically, UCIRC2's two IR cameras observe at wavelengths of 10$\mu$m and one at 12$\mu$m. A calibrated image in a single frequency band can be used to determine the temperature of an object of known emissivity ($\epsilon$), but cloud emissivity is highly variable and significantly less than 1. Thus, a multifrequency observation is required to break the degeneracy between $\epsilon$ and $T_{\rm c}$. For a single layer of clouds above an ocean of known surface temperature (and thus power, $P_{\rm E}$), one can estimate power on the detector, $P_{\rm tot}$ as,

\begin{equation}
    P_{\rm tot} = \epsilon P_{c}+(1-\epsilon)P_{E}
\end{equation}

Here, $P_{c}$ is the power of the cloud, from which $T_{c}$ and thus CTH, can be inferred. Other methods for reconstructing CTH can be found in \cite{Mario}.

\section{Design}

\subsection{IR Cameras}

UCIRC2 is outfitted with two $640\times480$ pixel Sierra Olympic Viento G uncooled IR cameras with 9mm lenses, focused at infinity. The cameras have a wide FoV, ($70^\circ \times 50^\circ$), such that the IR cameras have a significantly larger FoV than EUSO-SPB2's fluorescence telescope. When the payload is in ``night mode'', which occurs when the atmosphere is dark enough to allow for proper functioning of photodetection modules (PDMs), UCIRC takes a pair of pictures every minute. The wide field of view of the IR cameras  makes it possible to extrapolate the cloud conditions in the section of the atmosphere swept out by the PDM field of view in the time between pictures. 

The native spectral response of the cameras is 7.5 to 13.5 microns, but each camera is fitted with a different filter in order to facilitate the radiative CTH reconstruction methods. One of the cameras is fitted with a SPECTROGON bandpass light filter which transmits wavelengths between 11.5 and 12.9$\mu$m (denoted 12$\mu$m). The other camera is fitted with an Edmund Optics bandpass light filter that transmits wavelengths between 9.6 and 11.6$\mu$m (denoted 10$\mu$m). These bands are spaced to obtain brightness temperature data that facilitates both the Blackbody Power Ratio CTH reconstrction method and the Radiative Transfer Equation CTH reconstruction method. 

The cameras are powered via a power-over-ethernet (PoE) connection, which also enables communication with a Minnoboard Turbot, herein referred to as the UCIRC2 CPU. 

\subsection{Software}

The Sierra Olympic cameras can be operated via the Pleora eBUS Player software, a graphic user interface (GUI) which generates images from IR camera output. In order to automatically capture and store lossless images without a GUI, the eBUS Player will be redesigned using the eBUS software development kit (SDK), such that the raw ADC output from each microbolometer pixel is saved every 60 seconds. In order to improve the signal to noise ratio of the resulting data, a burst of images will be captured and added together. This ``sum image" is then compressed using bzip2, such that it requires $\lesssim 250$kB of storage on the UCIRC2 CPU. Since two sum images (one per camera) are captured every 60 seconds, UCIRC2 will generate approximately 0.5MB of data per minute. The dual core Atom processor on the MinnowBoard can process two threads concurrently, allowing for nearly simultaneous ($\Delta t \lesssim 10$ms) image capture. 

Each camera will be powered via ethernet which can be switched off using the GPIO pins of the MinnowBoard. Periodically power cycling the cameras allows them to reconnect with an ethernet hub which connects them to the UCIRC2 CPU. This power cycling can be carried out with GPIO pin commands. Namely, the GPIO signal will switch a pair of Potter Brumsfeld latching relays driven by a transistor current amplifier, thereby directing ON/OFF power signals to the cameras.

\subsection{Environment Control}

\begin{figure}
\begin{center}
\includegraphics [width=.95\textwidth]{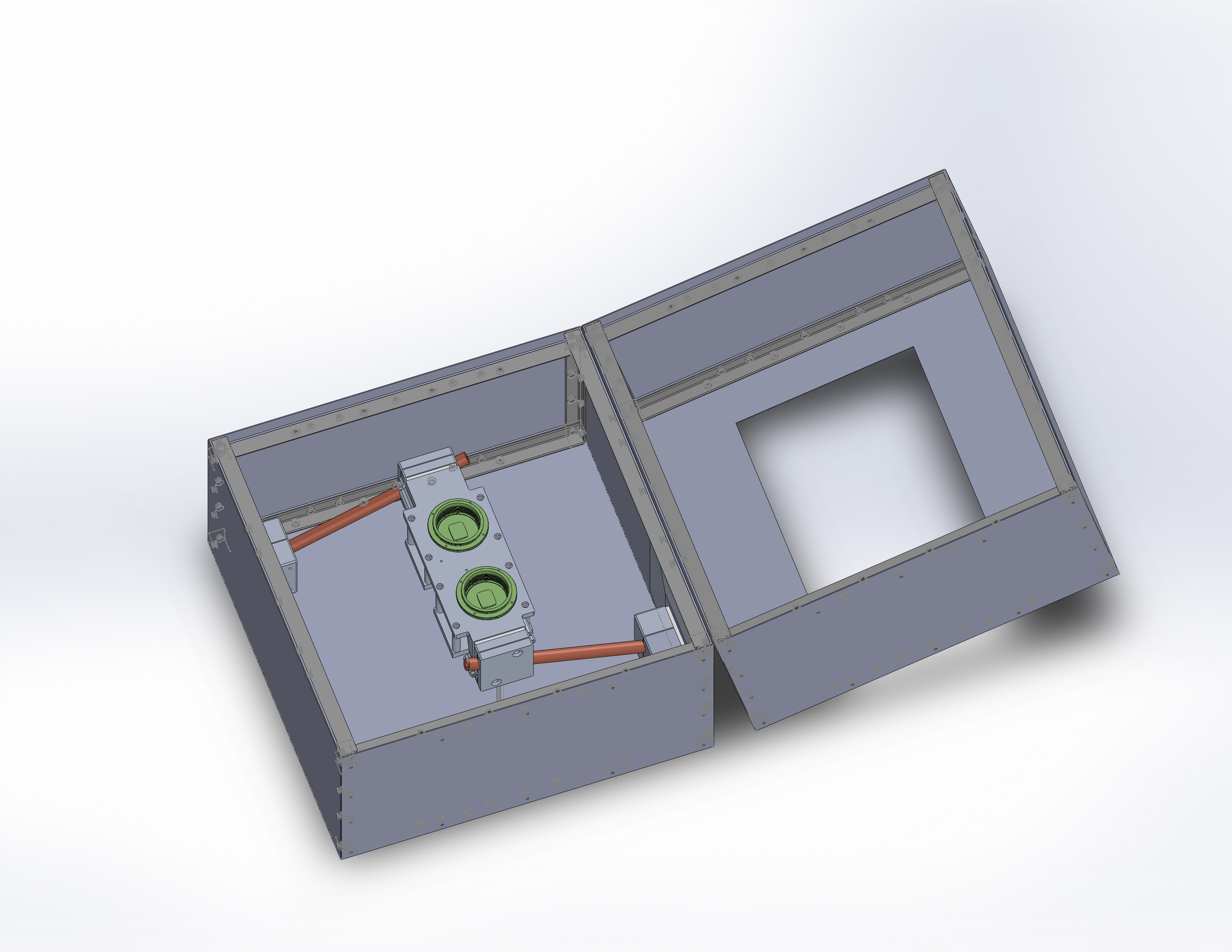}
\caption{Rendering of UCIRC2, with cameras (green circles) pointed toward the viewer and aluminum box open. 10.5 and 12 micron filters will be mounted in front of each camera. A system of peltier coolers (white rectangular prisms), heat pipes (copper-colored tubes), and resistive heaters (not shown) will maintain a steady camera temperature. The entire system will be enclosed in an aluminum box (grey structure) which will filled with insulating foam (not shown) and coated with emissive paint (not shown).}
\label{fig:schematic}
\end{center}
\end{figure}

\begin{figure}
    \centering
    \includegraphics[width=0.95\textwidth]{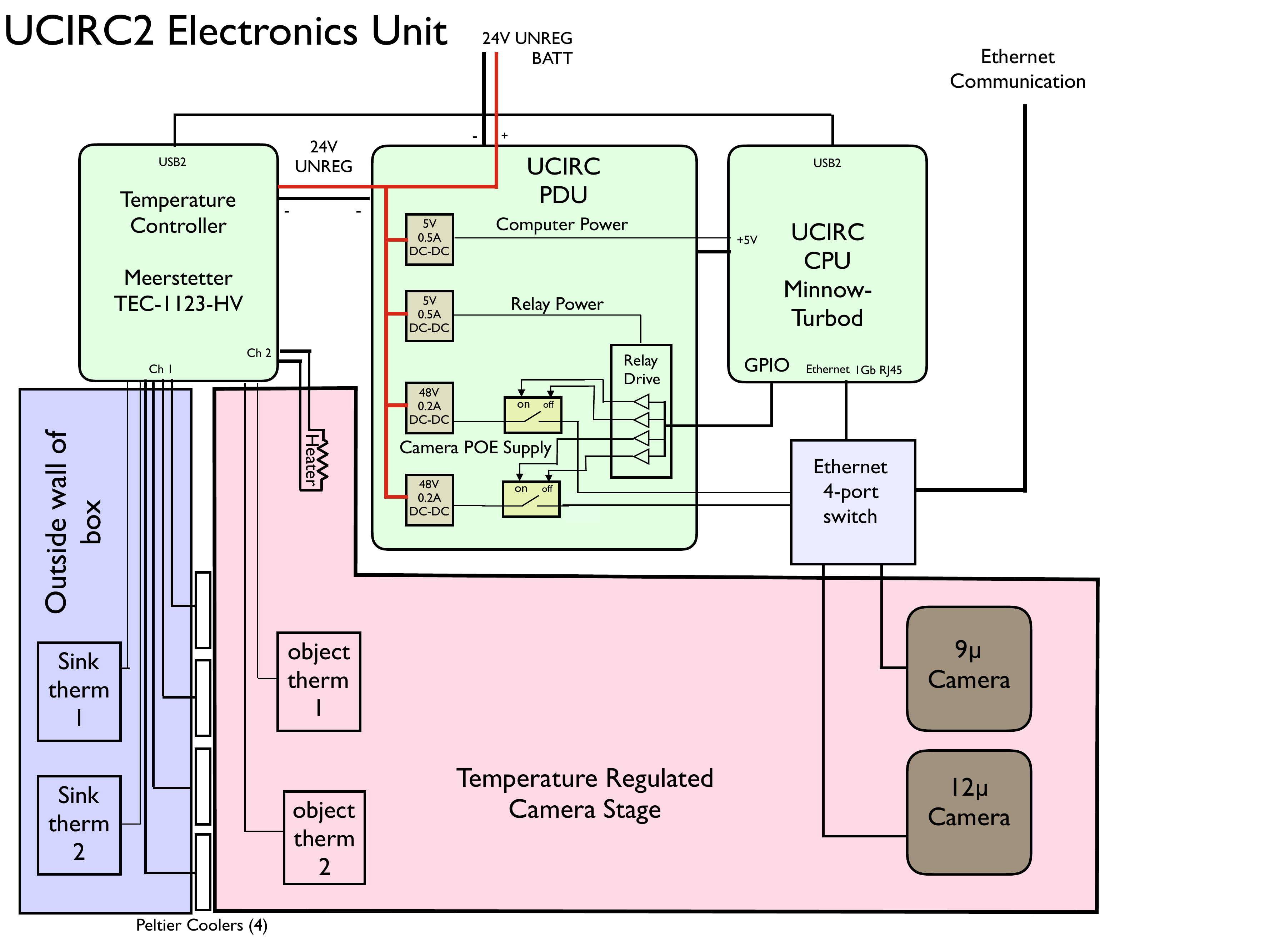}
    \caption{A block diagram of the UCIRC2 electronics and temperature management system. The two cameras communicate with the UCIRC2 CPU via ethernet using a two port hub. The CPU initializes image acquisition, compresses images, and stores them. The cameras are mounted on a temperature stabilized platform controlled via a thermal controller using Peltier coolers and a resistive heater. Unregulated 24V battery power is converted to 5V for the CPU and 48V for the cameras. Power to the cameras can be individually be switched on and off by the CPU. The electronics are passively cooled via heat pipes.}
    \label{fig:thermal}
\end{figure}

UCIRC2 is designed to operate in a high altitude ($\approx 33$km) environment during both daytime, when ambient temperatures reaches approximately 40C, and nighttime, when ambient temperatures reach approximately -40C. Temperature management is therefore a central design concern. In particular, the camera response is temperature dependent, meaning that camera temperature must be held approximately constant during operation (night mode). To maintain a stable temperature, the two cameras are housed in a $300$mm$\times$300mm$\times300$mm aluminum box coated with high emissivity flat white paint. This box is hinged and can be easily opened and closed to allow access to the cameras and electronics. Inside the box, the components of UCIRC2 are embedded in insulation foam. Most importantly, an active temperature management system consisting of heaters, Peltier coolers, heat pipes, and thermometers enables precise temperature monitoring and control (see Figures \ref{fig:schematic} and \ref{fig:thermal}).

The active heating and cooling system is controlled by a two-channel Meerstetter Engineering HV-1123 thermoelectric cooling and heating controller (TEC). Channel 1 of the TEC drives current to four Laird Technologies 56995-501 30mm$\times$30mm Peltier coolers. The second channel of the TEC is connected to a 10 $\Omega$ Vishay Dale resistive heater. The resistive heater is the primary mechanism for heating the camera stage and is used in combination with the Channel 1 Peltier system when heating is needed. The active temperature control system is designed to be most effective at heating because, in general, UCIRC2 will be actively collecting data during the nighttime, when the environment is cold. 

The temperature regulated camera stage is a machined aluminum plate, called the main stage, to which both IR cameras are thermally coupled. This stage is connected with good thermal contact to two Peltier coolers, which connect via two Enertron sintered powder wick copper heat pipes leading to two additional Peltier units at the outside panels. These heat pipes transfer the heat pumped by the Peltier units between the camera stage and the aluminum side walls. 

A second thermal stage supports the electronics boards (the CPU, the TEC, the USB hub and the power distribution board). This stage is cooled passively by heat pipes.

The TEC-1123-HV uses four temperature sensors. Two NTC thermistors with 5k$\Omega$ resistance at 20C measure the heat sink temperature, where two of the Peltier coolers contact the side walls. The other two sensors, Platinum Resistance Thermometers (PRTs) operated with a four wire readout, measure the temperature of the center and the edge of the camera thermal stage. The temperature measurement of the center of the camera thermal stage controls the TEC Proportional Integral Derivative controller (PID) in order to keep the main thermal stage at a constant, settable temperature. The second sensor is run as a monitor but does not control the PID loop. The set point temperature for the cameras can be modified by telemetery command, with daytime and nighttime operating temperatures chosen to minimize power consumption. 

\section{Testing and Calibration}

\begin{figure}[ht]
    \centering
    \includegraphics[width=0.9\textwidth]{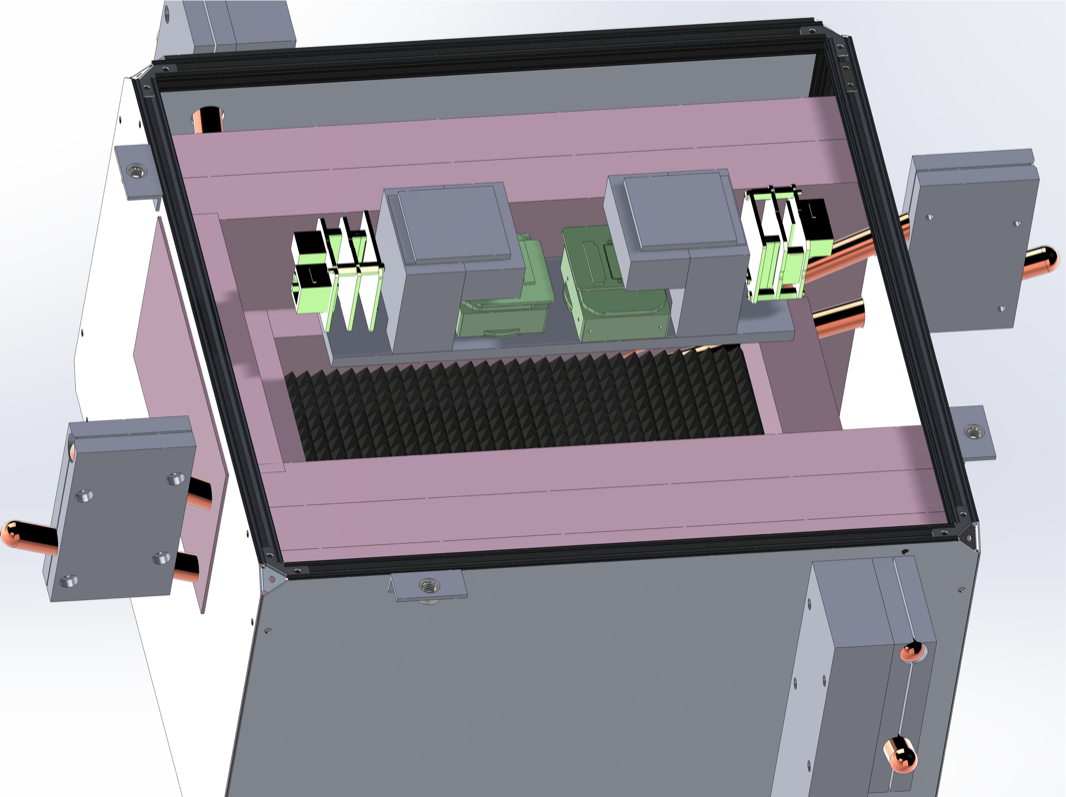}
    \caption{Rendering of a preliminary calibration setup, in which both cameras (green) are mounted above a calibrator. This calibrator consists of an aluminum box housing a temperature-controlled blackbody target (black spikes). As in UCIRC2, the temperature of the target is controlled by a system of peltier coolers, heat pipes, resistive heaters, and insulating foam (pink material). Once UCIRC2 has been constructed, the calibrator will be mounted onto the front of UCIRC2 in place of a lens cap, allowing for additional testing of the constructed system.}
    \label{fig:calibrator}
\end{figure}

To replicate the expected flight environment, UCIRC2 will be tested in a thermovac chamber pumped down to 0.3 mbar and cooled with liquid nitrogen. The temperature management system will be tested over all possible environmental temperatures to ensure that the cameras can be maintained within their operating temperature range. 

To calibrate the cameras, UCIRC2 will be positioned above a calibration target consisting of a highly emissive, temperature-controlled material (see Figure \ref{fig:calibrator}. By taking of images of the calibration target at multiple temperatures, this target will be used to perform a pixel by pixel calibration of each camera. Because the cameras' thermal response depends on their temperature, calibration images will also be taken at multiple camera temperatures. Before flight, the camera response will be tested under vacuum over all potential environmental and target temperatures.

\section{Acknowledgements}
UCIRC2 is supported by NASA Grant 80NSSC18K0246 and acknowledges previous work from the UCIRC1 team and the JEM-EUSO collaboration. Mehling and Friedlander acknowledge support from the Kavli Institute for Cosmological Physics.

\end{document}